\documentclass[prb,groupedaddress,twocolumn,showkeys,showpacs,floatfix]{revtex4}
\usepackage{epsfig}
\usepackage{color}
\usepackage{amssymb,amsmath}
\usepackage{slashbox}
\usepackage{makecell}
\usepackage{color}
\begin{document}
\title{Effect of picosecond magnetic pulse on dynamics of electron's subbands in
semiconductor bilayer nanowire}
\author{T. Chwiej}
\email{chwiej@fis.agh.edu.pl}
\affiliation{AGH University of Science and Technology, Faculty of Physics and Applied Computer
Science, al. A. Mickiewicza 30, 30-059 Cracow,
Poland}
\begin{abstract}
\noindent
We report on possibility of charge current generation in nanowire made of two tunnel coupled
one-dimensional electron waveguides by means of  single magnetic pulse lasting up to 20 ps.
Existence of interlayer tunnel coupling plays a crucial role in the effect described here as it
allows for hybridization of the wave functions localized in different layers which can be
dynamically modified by applying a time changeable in-plane magnetic field.
Results of time-dependent DFT calculations performed for a bilayer nanowire
confining many electrons show that the effect of such magnetic hybridization relies on tilting of
electrons' energy subbands, to the left or to the right, depending on a sign of time derivative of
oscillating magnetic field due to the Faraday law. Consequently, the tilted subbands become a source
of charge flow  along the wire. Strength of such magneto-induced current oscillations may achieve
even $0.6\mu\textrm{A}$ but it depends on duration of magnetic pulse as well as on
charge density confined in nanowire which has to be unequally distributed between both transport
layers to observe this effect.

\end{abstract}
\maketitle

\section{Introduction}

Single quantum wires (SQWr) are the objects in which the charge carriers are confined in two
dimensions but can move freely (ballistically) along the wire on the distance exceeding 
$10\mu\textrm{m}$.\cite{fischer2} Due to quantized motion of electrons in transverse
direction, their energies form continuous subbands, which are occupied up to the Fermi level and for
this reason an activation of subsequent subband is visible as a sudden upward step in conductance
measurements.\cite{thomas,smith2} Thus SQWr may constitute a basic building blocks in
vast majority of nanodevices designed for experiments probing the quantum phenomena in
transport measurements.\cite{wigner2,smith,hew,bielejec}

If two SQWr are aligned laterally\cite{eugster,wigner2} or
vertically,\cite{fischer2,fischer4,thomas} one over another, the electronic properties of such
double quantum wire system (DQWr) are remarkably modified due to their electrostatic and tunnel
coupling. For example, in vertically stacked DWQr system, the strength of tunnel coupling, besides
the actual geometry of nanostructure, i.e. the width of a barrier separating the wires, can be
selectively modified by applying external magnetic
field.\cite{thomas,lyo1,lyo2,fischer2,fischer4,chwiej_physb,chwiej_physe}
In particular the in-plane orientation of
magnetic field plays a crucial role in transport measurements. Fischer et all. in
work\cite{fischer2} by applying an in-plane magnetic field in longitudinal (along the
wire) and then
in lateral directions were able to identify single transport modes and the energy splittings between
subsequent subbands. Generally, there are two kind of effects  in which the  magnetic field
influences on a single electron's wave function in DQWr. First, it squeezes it in each wire
diminishing hence its tunneling motion which can be even completely turned off in strong fields
giving two separated transport channels.\cite{mourokh} Second, it hybridizes the ground state and
first excited state in vertical direction what modifies the energy
subbands.\cite{lyo1,lyo2,chwiej_physb,chwiej_physe} Such hybridization is activated by the
off-diagonal elements in Hamiltonian what transforms the subbands crossings in energy spectrum into
avoiding crossings called pseudogaps.\cite{lyo1,chwiej_physb} Occurrence of pseudogaps is visible as
sudden drop in conductance for increasing Fermi energy what can be realized by applying appropriate
voltages to the top and back gates.\cite{smith2,bertoni}
In recent years, attention of researchers mainly attract the many-body effects appearing in bilayer
nanosystems which can be examined in quantum transport measurements. To name some, these studies
focus on the formation of Wigner crystals,\cite{wigner,wigner2,hew} properties of composite fermions
in quantum Hall regime in bilayer systems\cite{composite_fermions}, Coulomb drag
currents\cite{wigner2,drag} or ferromagnetism which appearance in strictly one-dimensional systems
is forbidden due to Lieb and Mattis theorem but as shown by Wang et all. in
work\cite{ferromagnetism} can potentially be realized in DWQR due to the inter-wire tunneling.

Quite recently Chwiej has introduced in work [\cite{chwiej_accel}] a simple model describing an
interaction of single electron confined in vertical bilayer nanowire with a picosecond magnetic
pulse.\cite{auston,spin_dynamics,offresonant} There was shown that the fast oscillating magnetic
field can effectively change the motion energy of electron, provided that the magnetic field is
perpendicular to direction of electron transport and to direction of interlayer tunnel coupling. In
such case, electron can be magnetically accelerated in nanowire in a direction depending on
polarization of magnetic stimulus and initial division of charge between the coupled layers.

Present work is an extension of that idea to a many-electron case. Here, electrons are confined in
infinite nanowire that consists of two one-dimensional electron waveguides separated by thin
tunnel barrier. We assume the wire operates in the ballistic regime and consequently the electrons
are described by the Bloch states\cite{ihnatsenka} which form the subbands in energy spectrum
$E(k)$.
These subbands are sampled for discrete values of wave vector $k$ and the corresponding wave
functions
are involved in time-dependent DFT calculations.
Calculated energies of chosen Bloch states allow to reconstruct the actual shapes of energy
subbands when the considered nanosystem interacts with a single picosecond magnetic pulse.
Results show that energy subbands can be tilted for short period of
time due to interaction of electrons with oscillating magnetic field. That leads to
imbalance between the number of electrons having positive and negative wave vectors in vicinity of
the Fermi energy level. Such tilting of subbands directly generates a single charge current
oscillation flowing along the wire.
We show that amplitude of such magnetically induced current increases linearly with the duration of
magnetic pulse. Moreover, we notice that the charge confined in nanostructure should be unequally
distributed between two transport layers exactly as  in a single electron
problem.\cite{chwiej_accel} We indicate the last condition is crucial for considered effect to be
observable.

The paper is organized as follows. In Sec.\ref{Sec:theo} we first describe in detail
the structural properties of nanostructure we study, and then we present a DFT based numerical model
used in calculations of energy spectra in nanowire. Section \ref{Sec:res} is devoted to presentation
and discussion of numerical results while conclusions are given in Sec.\ref{Sec:con}.

\section{Theoretical model}
\label{Sec:theo}

\begin{figure}[htbp!]
\epsfig{file=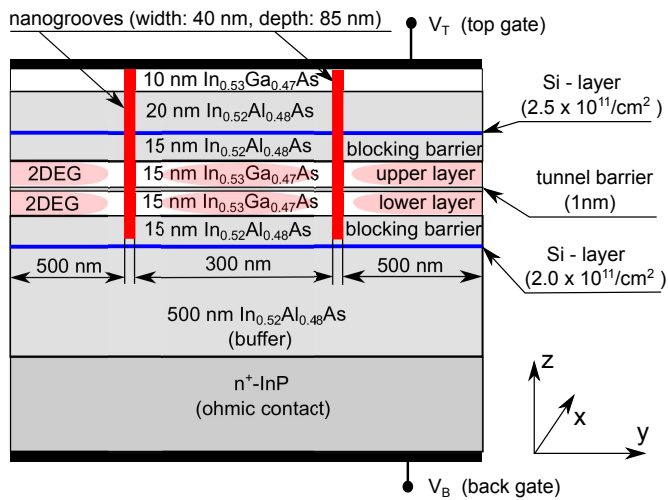,width=65mm,clip="6 5 322 431",angle=-90}
\caption{(Color online) Cross section of  bilayer nanowire considered in work. }
\label{Fig:model}
\end{figure}

In  the following considerations we assume  the electrons are confined in vertical
(z-axis) direction within two  $15\textrm{-nm-wide}$ quantum wells made of
$\textrm{In}_{0.53}\textrm{Ga}_{0.47}\textrm{As}$ which are separated by a
thin $1\textrm{-nm-wide}$ barrier and are surrounded from bottom and top by wide
$\textrm{In}_{0.52}\textrm{Al}_{0.48}\textrm{As}$ barriers  [see
Fig.\ref{Fig:model}]. The height of barriers equals $504\,\textrm{meV}$.
The electrons are provided by two $\delta\textrm{-doped}$ donors layers
localized below
and above the double-quantum-well (DQW) structure. Due to proximity of these positively ionized
doped layers ($15\,\textrm{nm}$ below and above the DQW), the conduction band in DQW is bending
toward the bottom and the top, in the lower and
in the upper quantum wells, respectively.\cite{fischer,chwiej_physe} Such enhancement of vertical
confinement within each
quantum well leads to formation of two, the lower and upper, transport layers, which are tunnel
coupled. The lateral confinement in y direction can be realized by application of
anodic oxidization of surface technique. By ploughing the surface with an atomic force microscopy
scanner, two parallel nanogrooves can be made,\cite{vgroove,vgroove2} along  x-axis in our case,
which separate the electrons localized in the middle part of nanostructure from the left and right
parts of the 2DEG [see Fig.\ref{Fig:model}]. This
central region forms actually the bilayer nanowire, in which the electrons can
move freely along the wire but their motion in transverse directions becomes quantized. 
The electrons fill the DQW structure up to the Fermi energy level, which value is fixed at
$E_{F}=0$ by two electron reservoirs (source and drain) attached to the ends of the lead.
However both, the top gate covering the surface of nanostructure and the back gate lying at the
bottom, allow changing the electron densities selectively in each layers by applying
appropriate voltages to the top ($V_{T}$) and back ($V_{B}$) gates as
has been shown in works by Fischer et al. [\cite{fischer2,fischer}] for 
 $\textrm{AlGaAs/GaAs}$  heterojunction based nanostructure.
 Here we use similar geometry for
$\textrm{In}_{0.52}\textrm{Al}_{0.48}\textrm{As}/\textrm{In}_{0.53}\textrm{Ga}_{0.47}\textrm{As}$
nanostructure as the electrons confined in DQW have lighter effective mass
($m_{InGaAs}^{*}=0.04$ versus $m_{GaAs}^{*}=0.067$) what enhances their tunneling rate and
speeds up the response of electron gas to the time variations of magnetic field.

Reaction of the conducting electrons to magnetic pulse and therefore the amplitude on
magneto-induced current depends on the magnitude of hybridization in vertical parts of their wave
functions.\cite{lyo1,fischer2,chwiej_physb} This, in turn, mainly depends on the interlayer tunnel
strength and therefore of
particular interest is determination of actual confining potential landscape. For this purpose we
employ the electrostatic model described in work [\cite{chwiej_physe}] which has been worked out
exactly for the geometry of nanodevice considered here.

We study the electronic structure of bilayer nanowire for Fermi energy
exceeding $10\,\textrm{meV}$ and since single electron kinetic energy dominates over the
electron-electron correlation energy,\cite{wigner2} we are justified in providing the further
analysis of electronic properties in language of density functionals.
Within DFT approximation  Hamiltonian of single electron is given by
\begin{equation}
 \widehat{h}=
 \frac{\left(\widehat{\pmb{p}}+e\pmb{A} \right)^2}{2m^{*}}
 \pm\frac{1}{2}g\mu_{b}B
 +V_{conf}(y,z)+V_{xc}^{\sigma}(y,z)+V_{H}(y,z),
 \label{Eq:ham}
\end{equation}
where $m^{*}=0.04$ is the conduction band effective mass in
$\textrm{In}_{0.53}\textrm{Ga}_{0.47}\textrm{As}$ quantum well,
$\widehat{\pmb{p}}=-i\hbar\nabla$ is a momentum operator of electron,
$V_{conf}$ denotes an external confining potential,
$V_{xc}^{\sigma}$ is an exchange-correlation potential calculated within
a local-spin-density-approximation, while $V_{H}$ is the Coulomb
part of electrostatic interaction obtained as solution of Poisson's equation.
The details of calculations of $V_{xc}$ and $V_{H}$ are given in work [\cite{chwiej_physe}].
Second term in Eq.\ref{Eq:ham} describes the contribution due to spin Zeeman effect with
Lande factor $g=-4$, while the $\pm$ signs correspond to electron spin being parallel
($\sigma=\uparrow$) and
antiparallel ($\sigma=\downarrow$) to magnetic field.
For vector potential $\pmb{A}$  we use a non-symmetric gauge
$\pmb{A}=\left(z'\,B_{y}-y'\,B_{z},0,0\right)$ where $z'$ and $y'$ are defined as  $z'=z-z_{0}$ and
$y'=y-y_{0}$ with $(y_{0},z_{0})$ being a point at the center of tunnel barrier at half
width of nanostructure shown in Fig.\ref{Fig:model}.
We assume $B_{z}=const$,  while $B_{y}$ changes
with time. The following expression defines the time characteristic of magnetic pulse used in
calculations
\begin{equation}
B_{y}(t)=1.3\,B_{y}^{max}\sin(\Omega_{y}\,t)\,\sin(\Omega_{y}\,t/2)\, \theta(t)\,
\theta(t_{imp}-t),
\label{Eq:pulse}
\end{equation}
where $\theta(t)$ is Heaviside step function, $t_{imp}$ denotes the length of magnetic pulse and
$\Omega_{y}=2\pi/t_{imp}$ its frequency.
For all results presented below the amplitude of magnetic pulse is $B_{y}^{max}=0.5\,\textrm{T}$.
This value can be easily reached in practical realization with repetition frequency exceeding
$100\,\textrm{kHz}$.\cite{spin_dynamics}
Shape of magnetic pulse defined in Eq.\ref{Eq:pulse} is depicted in Fig.\ref{Fig:iq}(a).
Due to the translational
invariance of the confining potential, the wave function of electron with spin $\sigma$ can be
written as a plane wave
\begin{equation}
\Psi_{n,k,\sigma}(\pmb{r},t)=\frac{1}{\sqrt{2\pi}}\varphi_{n,k,\sigma}(y,z,t)e^{ikx},
\label{Eq:psi}
\end{equation}
where $\varphi_{n,k,\sigma}(y,z,t)$ describes the part of wave function for transverse
direction,which generally can be time-dependent. In single particle picture involved here, the
electrons have well defined wave vectors $k$ and form the energy subbands which are denoted by
index $n$. The main aim of this paper is to show the dynamic response of these subbands to stimulus
in form of a picosecond magnetic pulse. By introducing the plane wave approximation in
Eq.\ref{Eq:psi} we assume that electrons move ballistically only along the wire, however they
can still be scattered in vertical and lateral directions due to the combined effect of
non-homogeneity in the confining potential and variations of $B_{y}(t)$ which contributing to the
magnetic force temporarily deflects the trajectories of electrons.[\cite{chwiej_accel}]

Calculations of $\varphi_{n,k,\sigma}(y,z)$ are performed on a rectangular spatial mesh of nodes in
y-z plane, that is, $y=i\cdot \Delta y$ and $z=j\cdot \Delta z$ for $\Delta y=2\,\textrm{nm}$ and
$\Delta
z=0.5\,\textrm{nm}$. Including the Peierls phase shift in kinetic operator for mesh in  x
direction\cite{chwiej_physe} ($\Delta x=\Delta y$) and then
averaging the Hamiltonian (Eq.\ref{Eq:ham}) over the x variable,
$\langle\widehat{H}\rangle=\langle e^{ikx}|H|e^{ik'x}\rangle=\widehat{h}\delta(k-k')$,
one gets the effective energy operator for the wave function $\varphi_{n,k,\sigma}(y,z)$
\begin{eqnarray}
\nonumber
\widehat{h}&=& \frac{\hbar^2}{m^{*}\Delta x^2}\left[1-cos\left(k\,\Delta x+ \frac{m^{*}\Delta
x}{\hbar}(z'\,\omega_{y}(t)-y'\,\omega_{z})
\right) \right]\\
&+&\frac{\widehat{p}_{y}^{2}+\widehat{p}_{z}^{2}}{2m^{*}}
+V_{tot}^{\sigma}.
\label{Eq:ham2}
\end{eqnarray}
In Eq.\ref{Eq:ham2}, $V_{tot}^{\sigma}$ is the sum of all potentials appearing in
Hamiltonian (\ref{Eq:ham}) and the Zeeman term.  The first kinetic term in Eq.\ref{Eq:ham2} depends
on wave vector $k$ but
also on $z$ and $y$ variables if the cyclotron frequencies $\omega_{y}(t)=eB_{y}(t)/m^{*}$ and
$\omega_{z}=eB_{z}/m^{*}$ do not vanish.

From Eq.\ref{Eq:ham2} appears that although the variations of $B_{y}(t)$ can not
change the canonical wave vector $k$, they may change both, the group velocity of electron
($v_{gr}$), assuming that $B_{z}=const$, and its motion energy which contributes to total energy
$E$, since these two quantities are
connected by formula
\begin{equation}
v_{gr}=\frac{1}{\hbar}\frac{\partial E}{\partial k}.
\label{Eq:vgr}
\end{equation}
Let us note that the wave function
$\varphi_{n,k,\sigma}(y,z,t)$ must be dependent on the wave vector's value
if $B_{z}\ne 0$ (and/or $B_{y}\ne 0$) what influences on the group velocity of electron
\begin{eqnarray}
\nonumber
v_{gr}(n,k,\sigma)&=&
\frac{1}{\hbar}\langle\varphi_{n,k,\sigma} |\partial_{k}\widehat{h}|\varphi_{n,k,\sigma} \rangle\\
&+&\frac{1}{\hbar}
2Re\left\{\langle\partial_{k}\varphi_{n,k,\sigma}|\widehat{h}|\varphi_{n,k,\sigma} \rangle
\right\}.
\end{eqnarray}
In such case, the effect of action of magnetic force on moving electrons relies on 
changing their localization in y direction in quantum wire what differentiates the
confinement energies of carriers for different wave vectors. It means that all subbands have no
longer simple parabolic shape.

To find out the dynamical subbands' responses to magnetic pulse, the $\Psi_{n,k,\sigma}$ states are
first prepared at $t=0$, and then, their transverse parts $\varphi_{n,k,\sigma}(y,z,0)$ evolve
in time according to time-dependent Schr\"odinger equation
$i\hbar\partial_{t}\varphi_{n,k,\sigma}=\widehat{h}\varphi_{n,k,\sigma}$.
Initial wave functions $\varphi_{n,k,\sigma}(y,z,0)$ are simply the eigenstates of Hamiltonian
(\ref{Eq:ham2}). They have been found during diagonalization of this energy operator on spatial mesh
at central part  of nanostructure shown in Fig.\ref{Fig:model} ($y=540\div
840\,\textrm{nm}$ and $z=506\div 557\,\textrm{nm}$). The time evolution of subbands is performed
with application of Magnus propagator\cite{magnus_propagator} for discrete values of $k$ taken at
$k_{l}=-1.3\cdot k_{F}+\Delta k\cdot l$ with $l=0,1,\ldots,n_{k}$ and  $\Delta k=2.6\cdot
k_{F}/n_{k}$, where $k_{F}$ is the
Fermi wave vector for first subband while $n_{k}=300$.
In other words $k_{F}$ is the maximal wave vector of electrons in nanosystem,
which should be determined separately for each initial state as it depends on voltages applied to
the top and back gates as well as on strength of $B_{z}$ (other parameters of calculations such as
dopants densities are fixed).  In calculations we used the time step $\Delta t=10^{-4}\,\textrm{ps}$
which guarantees stability of our numerical procedure and keeps errors on acceptable level.
Every $25$ time steps, the spin densities ($\rho_{\sigma}$) and total density
($\rho=\rho_{\uparrow}+\rho_{\downarrow}$) as well as the Hartree and
exchange-correlation potentials are recalculated. The spin densities are determined as follows
\begin{eqnarray}
\nonumber
 \rho_{\sigma}(y,z,t)&=&\frac{1}{2\pi}\sum_{n}\int\limits_{-\infty}^{\infty}dk
|\varphi_{n,k,\sigma}(y,z,t)|^{2}\,f_{E}\left(E_{k}\right)\\
&=&\frac{1}{2\pi}\sum_{n}\sum_{k_{l}}\Delta k
|\varphi_{n,k_{l},\sigma}(y,z,t)|^{2}\,f_{E}(E_{k_{l}}),
\end{eqnarray}
where n is the subband's index, $f_{E}$ is Fermi-Dirac distribution function and $E_{k_{l}}$ is
the energy of electron occupying $n\textrm{-th}$ subband with wave vector $k_{l}$.
In similar way we calculate the total charge current
\begin{equation}
 I(t)=\frac{1}{2\pi}\sum_{n}\sum_{\sigma}\int dx dy\int dk\, j_{n,k,\sigma}(t)
 f_{E}\left(E_{k}\right).
 \label{Eq:cur}
\end{equation}
Here the contribution to x-component of density current equals
$j_{n,k,\sigma}=-e\,v_{gr}(n,k,\sigma)$. Integrals in Eq.\ref{Eq:cur} are computed numerically and 
the temperature of electron gas used in calculations is $T=4.2\,\textrm{K}$.
In expression for $I(t)$ the  Fermi-Dirac distribution function is taken for
$t=0$. Thus, we explicitly assume that the backscattering resulting from intersubband
scattering, which potentially may lead to momentum relaxation, is absent in our model because we
work in the ballistic regime. However, an intersubband scattering without change of  wave
vector $k$ can still occur since it corresponds to mixing of two subbands by e.g. variations of
$B_{y}$ what locally influences on energy subbands' dispersions\cite{chwiej_physb,chwiej_physe}
and according to formula (\ref{Eq:vgr}) on group velocities.

\section{Results}
\label{Sec:res}

\begin{figure}[htbp!]
\epsfig{file=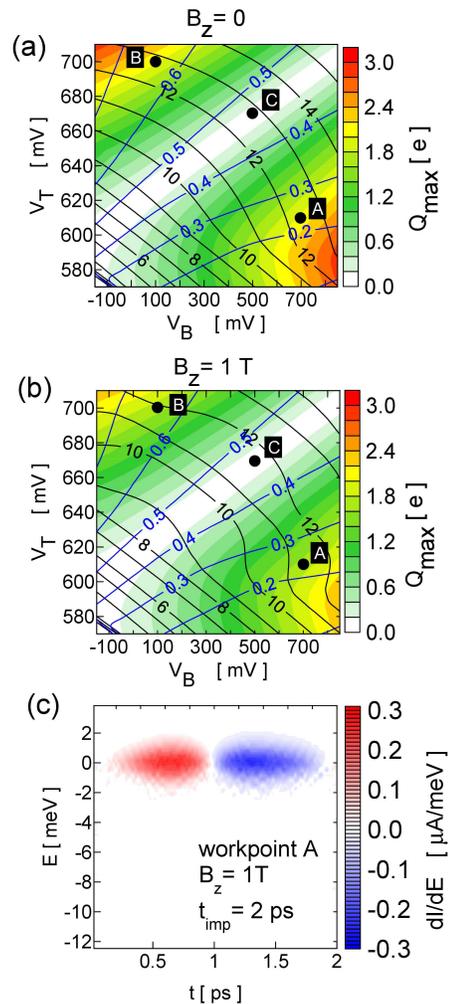,width=60mm,clip="0 0 375 842",angle=0}
\caption{(Color online) Amplitudes of charge that flows through
a nanowire for $B_{z}=0$ (a) and $B_{z}=1\,\textrm{T}$ (b) for $t_{imp}=2\,\textrm{ps}$.
(c) Energy resolved contributions to current for workpoint
A [as marked in (b)]. The black and blue contours in (a) and (b) show the Fermi
energy and the fraction of total charge confined in the upper layer, respectively.
}
\label{Fig:scan}
\end{figure}

\begin{figure}[htbp!]
\epsfig{file=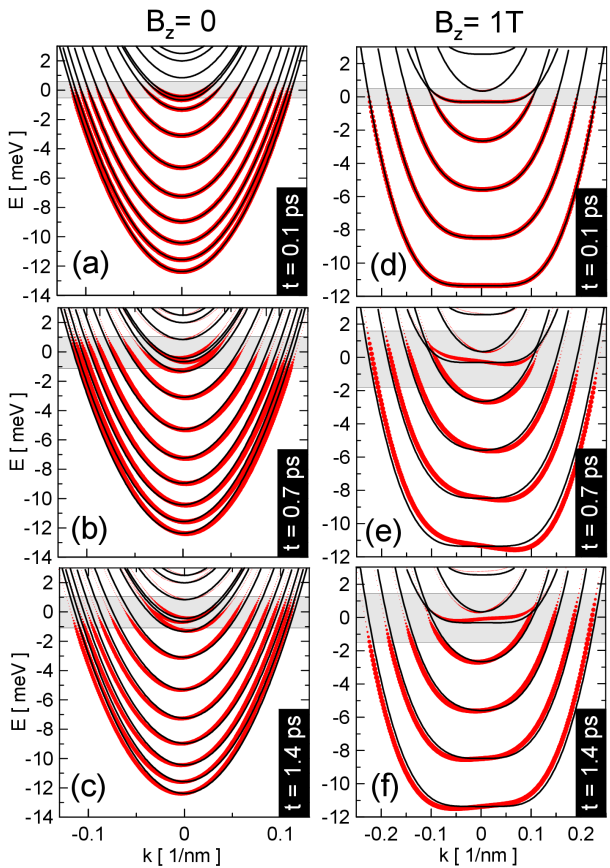,width=85mm,clip="0 235 436 842",angle=0}
\caption{(Color online) Snapshots of the lowest energy subbands for 
$B_{z}=0$ (left column)  and $B_{z}=1\,\textrm{T}$ (right column).
Subbands at $t=0$ are black, while these marked by the red dots were saved at time instants
displayed in the right-bottom of each chart. Grey horizontal strip in (a)-(f) shows approximately
the range of energy in which the subbands give contributions to the current. Size of
each red dot is proportional to occupation probability of particular state at $t=0$ calculated from
Fermi-Dirac function according to Eq.\ref{Eq:cur}.
}
\label{Fig:subbands}
\end{figure}

Figures \ref{Fig:scan}(a) and \ref{Fig:scan}(b) display the amplitudes of charge,
\begin{equation}
Q_{max}=\max\limits_{0<\tau< t_{imp}} \left|\int_{0}^{\tau} I(t) dt\right|,
\end{equation}
that flows through the nanowire when the electron gas confined in bilayer
nanostructure interacts with magnetic pulse of $2\textrm{-ps}$ duration.
Let us note that $Q_{max}$ changes qualitatively in the same way for $B_{z}=0$ and
$B_{z}=1\,\textrm{T}$. Its value increases when division of charge density between
the upper and lower layers is far from equilibrium (marked as white stripes). Then, the majority
of electron density is localized in one  layer what can be deduced from blue contours.
Maxima of
$Q_{max}$ are localized in top-left and bottom-right corners of Figs. \ref{Fig:scan}(a) and
\ref{Fig:scan}(b) showing thus strong dependence on gates biasing.
This fact immediately implies that appropriate selection of $V_{T}$ and $V_{B}$ voltages
shall enable one to choose the layer which holds the current and hence the direction of
charge flow.[\cite{chwiej_accel}] Application of static vertical component of
magnetic field ($B_{z}=1\,\textrm{T}$) diminishes $Q_{max}$ considerably.
$Q_{max}$ may completely disappear, provided that, the upper layer confines  slightly less
charge than the lower one what can be deduced from Figs. \ref{Fig:scan}(a) and \ref{Fig:scan}(b).
In such case, the current is still induced, but it flows in opposite directions in upper and
lower layers so both components cancel each other[\cite{chwiej_accel}] until separate gates are
attached to the upper and to the lower layer as shown by Bielejec et al. in work [\cite{bielejec}].

To get deeper insight into the dynamics of the electron subbands driven by time-varying
magnetic field, the results for three arbitrarily chosen workpoints marked in Figs.
\ref{Fig:scan}(a) and \ref{Fig:scan}(b) are analyzed in detail below.
They are defined by pair of ($V_{B},V_{T}$) voltages which are given in Tab.\ref{Table:data}.

\begin{table}[htbp]
\begin{ruledtabular}
\caption{Top gate $(V_{T})$ and back gate $(V_{B})$ voltages  in workpoints A, B and
C with corresponding sheet densities in the lower ($\rho_{low}$) and upper ($\rho_{up}$) layers.}
\label{Table:data}
\begin{tabular}{|c|c|c|c|c|c|}
workpoint &
\makecell[c]{$V_{B}$\\ $[\textrm{meV}]$} &
\makecell[c]{$V_{T}$\\ $[\textrm{meV}]$} &
\makecell[c]{$B_{z}$\\ $\left[\textrm{T}\right]$}&
\makecell[c]{$\rho_{up}$\\ $\left[10^{11}/\textrm{cm}^{2}\right]$}&
\makecell[c]{$\rho_{low}$\\ $\left[10^{11}/\textrm{cm}^{2}\right]$}\\\hline
A & $700$ & $610$ &$0;\, 1$ &  $0.56$ & $1.64$ \\\hline
B & $100$ & $700$ &$0;\, 1$ &  $1.64$ & $0.96$ \\\hline
C & $500$ & $670$ &$0;\, 1$ &  $1.30$ & $1.45$
\end{tabular}
\end{ruledtabular}
\end{table}

Fig.\ref{Fig:scan}(c) shows the energy resolved contributions to 
current in workpoint A with $B_{z}=1\,\textrm{T}$ and $t_{imp}=2\,\textrm{ps}$.
One can easily notice that the current is generated in vicinity of $E_{F}$ only indicating an
imbalance introduced to subbands by $B_{y}(t)$. At first, when
the polarization of magnetic pulse is positive, the current flows to the right and then it
disappears when polarization  is inverted. Afterwards the current starts flowing to the left.
Due to symmetry of the magnetic pulse defined in Eq.\ref{Eq:pulse} the amount of charge that was
initially shifted to the right and then to the left is almost identical.
Corresponding energy subbands saved at three time instants $t=0.1,\, 0.7\, \textrm{and}\,
1.4\,\textrm{ps}$ for the workpoint A, $t_{imp}=2\,\textrm{ps}$, $B_{z}=0$ and $B_{z}=1\,\textrm{T}$
are shown in Fig.\ref{Fig:subbands}. At $t=0.1\,\textrm{ps}$, when $B_{y}$ begins to grow,
subbands (red dots) do not differ much from the initial ones (black lines) and the
range of energy in which the Fermi-Dirac distribution function changes significantly is
about $2k_{B}T$ [see the narrow horizontal grey strip in Figs. \ref{Fig:subbands}(a) and
\ref{Fig:subbands}(d)].  At $t=0.7\,\textrm{ps}$, the influence of magnetic pulse on
subbands becomes noticeable, the branches with $k>0$ are lowered on energy scale
while these with  $k<0$ are lifted up widening hence the grey strip beyond an initial limit of
$2k_{B}T$. This remark, however, does not  concern the $8\textrm{-th}$, $10\textrm{-th}$ and
$11\textrm{-th}$ subbands for $B_{z}=0$. In these subbands the electrons occupy first excited state
in vertical direction and therefore they are localized in different layer (the upper one) than these
occupying the ground state in z (the lower layer). Such significant momentary tilt
of subbands on energy scale has to generate the current flow in nanowire. According to
Eq.\ref{Eq:cur} contributions to currents from the left ($k<0$) and right ($k>0$) moving electrons
do not cancel
mutually on a short time scale introduced by magnetic pulse.
Although, the energy shift between the states with $k>0$ and $k<0$ which belong to the same subband
is larger for $B_{z}=1\,\textrm{T}$ rather than for $B_{z}=0$, the intensity of current and
resulting $Q_{max}$ value is larger in the latter case [compare the scales in Figs.
\ref{Fig:scan}(a) and \ref{Fig:scan}(b)].
It results from the fact, that for $B_{z}=0$ there is $11$ active subbands near
the Fermi level which contribute to total current in workpoint A whereas for  $B_{z}=1\,\textrm{T}$
there is only $6$ of them. One must keep in mind however, that for $B_{z}=1\,\textrm{T}$ the slopes
of subbands at Fermi level are noticeably larger than for $B_{z}=0$ what directly influences on
current because of group velocity, given in Eq.\ref{Eq:vgr}, what in turn partially diminishes the
disproportion in current resulting from a large difference in number of active subbands.
At $t=1.4\,\textrm{ps}$ polarization of magnetic pulse is reversed and for this reason 
subbands are tilted in opposite direction [cf. Figs. \ref{Fig:subbands}(c) and
\ref{Fig:subbands}(f)]
what obviously reverses the direction of current flow in nanowire [see Fig.\ref{Fig:scan}(c)].

\begin{figure}[htbp!]
\epsfig{file=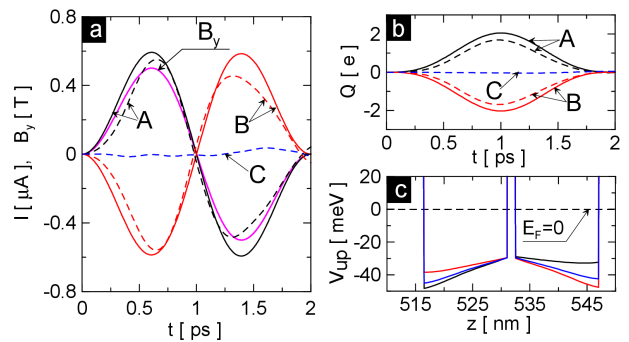,width=85mm,clip="0 568 476 842",angle=0}
\caption{(Color online) Time  dependences of current (a) and charge (b)
induced by $2\textrm{-ps-long}$ magnetic pulse for workpoints: A (black), B (red) and C (blue).
In (a) there is also displayed the magnetic pulse (pink) defined in Eq.\ref{Eq:pulse}.
(c) The profiles of the confining potential for spin-up electrons for workpoints A,
B and C. In (a) and (b) results obtained for $B_{z}=0$ and $B_{z}=1\,\textrm{T}$ are marked
with solid and dashed lines, respectively, while in (c) colors have the same meanings as in (a) and
(b). 
}
\label{Fig:iq}
\end{figure}

\begin{figure}[htbp!]
\epsfig{file=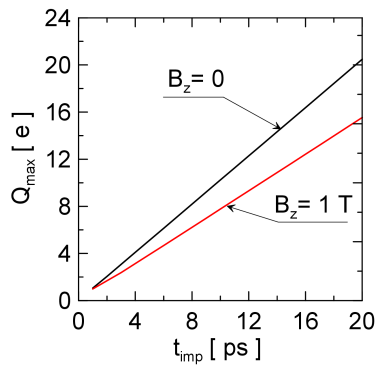,width=50mm,clip="0 472 377 842",angle=0}
\caption{(Color online) Dependence of $Q_{max}$ on magnetic pulse duration.
Results obtained for workpoint A.
}
\label{Fig:eff}
\end{figure}

The time characteristics of current generated for workpoints A, B and C are shown in
Fig.\ref{Fig:iq}(a). Even for short magnetic pulse ($t_{imp}=2\,\textrm{ps}$) its amplitude reaches
$0.6\,\mu\textrm{A}$ what makes its measurements experimentally feasible.
One can notice in this figure, that the current pulses for workpoints A and B resemble very much the
shape of the magnetic pulse [pink in Fig.\ref{Fig:iq}(a)] as they change their polarization exactly
at $t=1\,\textrm{ps}$. That
means that the mass inertia of electron density does not influence on the dynamics of subbands. The
time
characteristics of current and charge flow [see $Q(t)$ in Fig.\ref{Fig:iq}(b)] can be to some
extent modified by perpendicular magnetic field. For $B_{z}=1\,\textrm{T}$ the amplitudes of both
quantities have slightly lower amplitudes in comparison to results obtained for
$B_{z}=0$. Moreover they start growing with a certain delay what indicates
on influence of magnetic forces. This issue will be analyzed in detail further in text.
The momentary direction of current flow depends on whether
the major part of charge density is localized in upper layer or in the lower one.
Figure \ref{Fig:iq}(c) shows the vertical profile of the confining potential in the
center of nanowire. In workpoint A (black line) the majority of density is localized in
the lower deeper layer, while in workpoint B the upper layer is deeper. For this reason,
polarizations of current at these workpoints are opposite [see Fig.\ref{Fig:iq}(a)]. In third case,
in workpoint C, both layers confine similar amount of charge, and hence the electrons confined in
different layers are pushed in opposite directions\cite{chwiej_accel} giving thus no current flow
[see
Figs. \ref{Fig:iq}(a) and \ref{Fig:iq}(b)].

The amount of charge $Q_{max}$ carried by single current oscillation depends on the length of
magnetic pulse. As one may notice in Fig.\ref{Fig:eff}, dependence  $Q_{max}$ on $t_{imp}$ is
strictly linear but its slope decreases for $B_{z}\ne 0$. Thus, by tuning the values of parameters
such as $V_{B}$, $V_{T}$, $B_{z}$, $t_{imp}$ one can carry, forward and backward, precisely
determined amount of charge in bilayer nanowire.

\begin{figure}[htbp!]
\epsfig{file=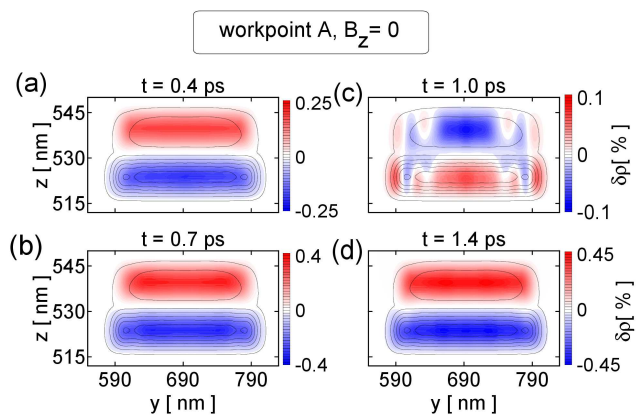,width=85mm,clip="0 462 595 842",angle=0}
\caption{(Color online) Snapshots of relative changes in electron density confined in
nanowire in workpoint A for $B_{z}=0$ and $t_{imp}=2\,\textrm{ps}$.
Black thin lines are the contours of initial (unperturbed) density for $t=0$, while the red and blue
colors mark the regions of increased and decreased density for $t>0$.
In (a)-(d) the time instants are displayed on top.
}
\label{Fig:roa0}
\end{figure}

Now let us  analyze the dynamics of intralayer and interlayer charge flow induced by magnetic pulse.
Figure \ref{Fig:roa0} shows the relative changes in  spatial distribution of density
in workpoint A for  $t=0.4,\, 0.7,\, 1.0\,\,\textrm{and}\, 1.4\,\textrm{ps}$ and $B_{z}=0$.
This quantity is defined as
\begin{equation}
\delta\rho(y,z,t)=\frac{\rho(y,z,t)-\rho(y,z,0)}{\rho_{max}}
\label{Eq:delta}
\end{equation}
where $\rho_{max}$ is the maximum of $\rho(y,z,0)$.
 At $t=0.4\,\textrm{ps}$ when the magnetic field $B_{y}$ is on its growing slope, a small part of
density is carried from the deeper
lower layer to the upper  one. Note that the amount of density is evenly
decreased in lower layer and evenly increased in the upper one. This process is continued until
$t=0.7\,\textrm{ps}$ [compare scales in Figs. \ref{Fig:roa0}(a) and \ref{Fig:roa0}(b)] and
afterwards an excess density comes back to lower layer what shows Fig.\ref{Fig:roa0}(c) for
$t=1.0\,\textrm{ps}$. Next, although polarization of $B_{y}$ and generated current are reversed
[see Fig.\ref{Fig:iq}(a)], part of the charge density flows again homogeneously from the lower layer
to the upper one what is shown in Fig.\ref{Fig:roa0}(d).
The reasons of this homogeneous charge flow visible in Figs. \ref{Fig:roa0}(a), \ref{Fig:roa0}(b)
and \ref{Fig:roa0}(d) are as follows.
For $B_{z}=0$, the dependence of $\varphi_{n,k,\sigma}$ on wave vector $k$ can be
neglected and the whole energy subband can be described by single wave function. $B_{y}(t)$
couples then both layers what hybridizes the ground state and the first
excited state in vertical direction but simultaneously it leaves the lateral excitations
(y direction) in $\varphi_{n,k,\sigma}$ unchanged. In other words, the wave function shape in
this direction
and the resulting lateral spatial distribution of charge density are preserved since both layers
have comparable widths. This picture is valid only if the interlayer charge flow is large as it is
shown in Figs. \ref{Fig:roa0}(a), \ref{Fig:roa0}(b) and \ref{Fig:roa0}(d).
Then the contributions from
the slowly oscillating in lateral direction lower subbands are significantly larger than
these being provided by strongly oscillating subbands activated at higher energies  what results
from much larger imbalance between $k>0$  and $k<0$ branches in
lowest energy subbands [see Figs. \ref{Fig:subbands}(b) and \ref{Fig:subbands}(c)].
If polarization of magnetic field is reversed what takes place at $t=1.0\,\textrm{ps}$,
contributions from all active subbands become comparable and as one may notice in
Fig.\ref{Fig:roa0}(c), oscillations in charge density occur near the edges of nanowire.

\begin{figure}[htbp!]
\epsfig{file=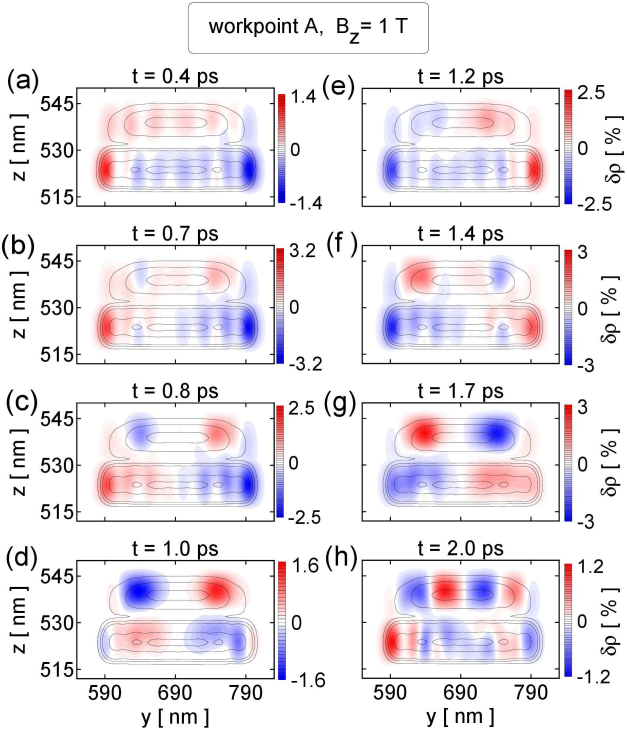,width=85mm,clip="0 126 595 842",angle=0}
\caption{(Color online) Snapshots of the relative changes in total electron density distribution in
nanowire in workpoint A and for  $B_{z}=1\,\textrm{T}$ and $t_{imp}=2\,\textrm{ps}$. Other markings
are the same as in Fig.\ref{Fig:roa0}.
}
\label{Fig:roa1}
\end{figure}

The mechanism of intralayer and interlayer charge redistribution driven by magnetic pulse is
modified in presence of perpendicular magnetic field. In Fig.\ref{Fig:roa1}, which shows
$\delta\rho$ for $B_{z}=1\,\textrm{T}$, we see that
 the interlayer charge flow is no longer homogeneous. First, the three lowest subbands for
$B_{z}=1\,\textrm{T}$ displayed in Figs.\ref{Fig:subbands}(d)-(f) have flat bottoms and their
energies strongly grow for large $k$ values what indicates formation of the edge states.
Electrons obey then the Lorentz force which pushes  electrons with $k>0$ and $k<0$ to the right
and to the left edge, respectively.

When $B_{y}$ increases, from the Faraday's law, $\nabla\times\pmb{E}=-\partial_{t} B_{y}$,
appears that the x-component of electric field induced in lower deeper layer accelerates the
electrons with $k<0$ and decelerates these with $k>0$.
Electrons with $k<0$ are hence stronger pushed to the left edge what increases their energies,
while the energies of electrons localized at the right edge ($k>0$) are decreased since these are
pushed towards the center of quantum well [see energy subbands in Fig.\ref{Fig:subbands}(e)].
As a result, a small fraction of charge is carried from the right edge to the left one
in lower layer and simultaneously from the lower deeper layer to the upper shallower one for
$t<0.5\,\textrm{ps}$ [Fig.\ref{Fig:roa1}(a)]. 
Next, for the time interval $t\approx 0.5\div 1.5\,\textrm{ps}$ the direction of induced electric
field is reversed due to negative value of $\partial_{t} B_{y}$. For this reason, the excess charge
localized in lower layer near its left edge  is continuously  carried to the right side, whereas the
charge confined in upper layer flows in opposite direction but with some time delay [see Figs.
\ref{Fig:roa1}(b)-\ref{Fig:roa1}(f)].
This tendency holds until  $t\approx 1.5\,\textrm{ps}$ when induced electric
field changes its direction again. That significantly diminishes the amplitude of charge
oscillations in lower layer especially near the left and right edges [cf. Figs. \ref{Fig:roa1}(f)
and \ref{Fig:roa1}(g)]. On the other hand it influences on charge oscillations in
upper layer with some delay  as these  have larger amplitude at $t=1.7\,\textrm{ps}$ rather than
those obtained for $t=1.4\,\textrm{ps}$.
Finally, when magnetic pulse vanishes for $t=2\,\textrm{ps}$ the amplitude of charge
oscillations in both layers are reduced but they are still visible [Fig.\ref{Fig:roa1}(h)].
Then, the charge distribution in lower layer resembles that obtained for $t=0.4\,\textrm{ps}$
because the directions of electric field induced in layers at the beginning and at end of the
magnetic pulse are the same since $\partial_{t}B_{y}>0$. The density oscillations in upper layer
are still distinct but they are two times frequent now what indicates the energy subbands lying
higher on energy scale are more involved.

\begin{figure}[htbp!]
\epsfig{file=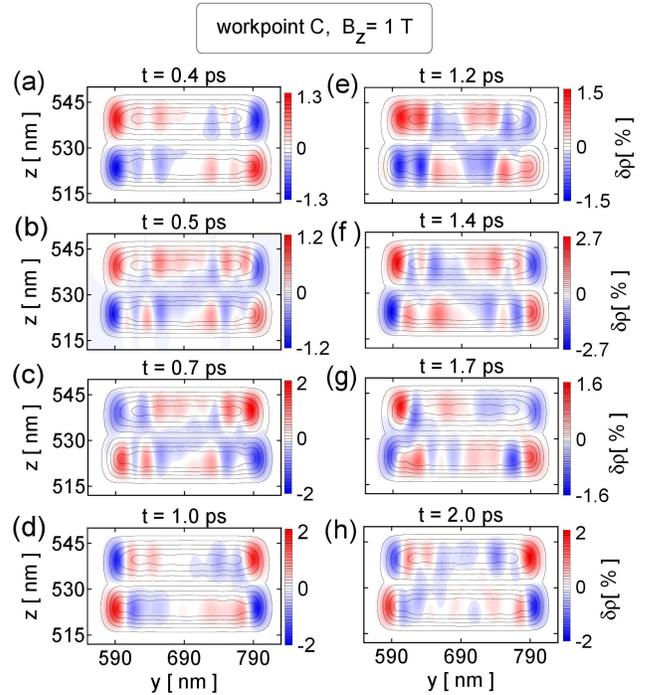,width=85mm,clip="0 316 595 842",angle=0}
\caption{(Color online) Time snapshots of $\delta\rho$ in workpoint C for
$B_{z}=1\,\textrm{T}$ and $t_{imp}=2\,\textrm{ps}$. Other markings are the same
as in Fig.\ref{Fig:roa0}.
}
\label{Fig:roc1}
\end{figure}

The time variations of $\delta\rho$ in workpoint C for $B_{z}=1\,\textrm{T}$ are presented
in Fig.\ref{Fig:roc1}. Since the upper and lower layers confine now $47\%$ and $53\%$ [see data in
Tab.\ref{Table:data}] of
charge, respectively, both layers play thus equivalent role in electron transport but their
contributions to the current cancel each other. In other words, the magnetic pulse can not generate
the charge flow between contacts attached to both ends of nanowire until
these are independently connected with upper and lower layers as it was shown in work by Bielejec et
al.\cite{bielejec} However, it may induce
local currents flowing in transverse directions. Snapshots of corresponding charge oscillations are
displayed in Fig.\ref{Fig:roc1}. Surprisingly, the effect of magnetic pulse on charge density in
lower layer at
$t=0.4\,\textrm{ps}$ is completely different than that observed in workpoint A [cf. Figs.
\ref{Fig:roa1}(a) and Figs. \ref{Fig:roa1}(b)]. Namely, the electron density confined in lower layer
is decreased at the left edge and increased on its right side 
[Fig.\ref{Fig:roc1}(a)] while the pattern of charge oscillations in upper layer is inverted.
Note, that for $t<0.5\,\textrm{ps}$ and $B_{z}=1\,\textrm{T}$, the rotational electric
field induced by magnetic pulse accelerates the electrons localized in lower layer and decelerates
these confined in upper layer, provided that considered electrons move near the left edge
with $k<0$.
However, besides the change of electrons' group velocities, magnetic pulse also bends their
trajectories in vertical direction changing their tunneling motion. The magnitude  of vertical
component of magnetic force, besides the strength of magnetic pulse, depends also on the group
velocity of electron. Therefore, the magnetic force enhances  the charge flow towards the upper
layer where the charge is accumulated but hinders the charge flow towards the lower layer where it
becomes depleted. At the right edge, the direction of charge accumulation is reversed due to
symmetry of the confining potential.

In workpoint C both layers have almost identical spatial sizes what implies that two subbands
which are defined by the same excitation mode in lateral direction but differ in vertical excitation
can be effectively hybridized by $B_{y}$.\cite{moroukh,fischer} That is true, provided that $B_{z}$
is not strong, otherwise the lateral component of magnetic force may diminish hybridization since
the wave functions' maxima of involved subbands do not coincide.
In workpoint C however, the charge can easily flow in vertical direction for $B_{z}=1\,\textrm{T}$
independently on y-coordinate of an electron, whereas at workpoint A the charge flow from the lower
layer to the upper one is blocked at edges by the potential barrier leading thus to its
accumulation in lower layer. For this reason the maxima of $\delta\rho$ visible in
Fig.\ref{Fig:roc1} for $t<0.5\,\textrm{ps}$ ($\partial_{t}B_{y}>0$) are localized in upper layer
while for $t=0.5\div 1.0\,\textrm{ps}$ ($\partial_{t}B_{y}<0$) in the lower one. Then, keeping in
mind that the time derivative of $B_{y}$ has negative values only for $t\approx 0.5\div
1.5\,\textrm{ps}$, the distinct
maximum appearing at the left edge in upper layer for longer time period, i.e. 
$t=1.7\,\textrm{ps}$ [see Fig.\ref{Fig:roc1}(g)], unambiguously indicates on time delay in charge
response to magnetic stimulus, likely due to a mass inertia of electron density.

\section{Conclusions}
\label{Sec:con}
In conclusion, the dynamics of energy subbands for electrons confined in bilayer nanowire was
theoretically studied.
It was shown that the time changeable magnetic field, which is perpendicular to the
directions of electron transport and interlayer tunnel coupling, is able 
to change the shapes of energy subbands for a short period of time
 approaching $2\,\textrm{ps}$. Due to the magnetic stimulus, the left and right parts of subbands
can be raised as well
as lowered on energy scale depending on the sign of $\partial_{t}B_{y}$ and on division of total
charge between two layers which has to be unequal. In such case, the momentary numbers of occupied
states for $k>0$ and for $k<0$ become different what in turn induces the 
current flow
along the wire. Shape and duration of such current pulse very well resemble that of
magnetic stimulus, while its amplitude may reach
$0.6\,\mu\textrm{A}$, what makes its experimental confirmation feasible. Actual current intensity
depends however on some factors such as the geometry of nanowire, density of dopants , strength and
duration of magnetic pulse as well as on disproportion in amounts of charge confined in the lower
and upper layers. The last factor can be easily modified by tuning the voltages applied to the top
and back gates. We hope the results presented in this work will encourage experimentalists to
perform measurements of magnetoinduced current for the nanodevice of the same or similar
construction.

\section*{Acknowledgements}
The work was financed by Polish Ministry of Science and Higher Education (MNiSW)
and was supported in part by PLGrid Infrastructure.
\section*{References}
\bibliography{lit_current}
\end{document}